\documentclass[12pt,epsf]{article}
\usepackage{graphicx}
\usepackage{epsfig}
\usepackage{amsmath}
\usepackage{subfigure}
\usepackage{slashbox}
\begin{document}

\def\a{\alpha}
\def\b{\beta}
\def\c{\varepsilon}
\def\d{\delta}
\def\e{\epsilon}
\def\f{\phi}
\def\g{\gamma}
\def\h{\theta}
\def\k{\kappa}
\def\l{\lambda}
\def\m{\mu}
\def\n{\nu}
\def\p{\psi}
\def\q{\partial}
\def\r{\rho}
\def\s{\sigma}
\def\t{\tau}
\def\u{\upsilon}
\def\v{\varphi}
\def\w{\omega}
\def\x{\xi}
\def\y{\eta}
\def\z{\zeta}
\def\D{\Delta}
\def\G{\Gamma}
\def\H{\Theta}
\def\L{\Lambda}
\def\F{\Phi}
\def\P{\Psi}
\def\S{\Sigma}

\def\o{\over}
\def\beq{\begin{eqnarray}}
\def\eeq{\end{eqnarray}}
\newcommand{\lsim}{\raisebox{0.6mm}{$\, <$} \hspace{-3.1mm}\raisebox{-1.5mm}{\em$\sim\,$}}
\newcommand{\gsim}{\raisebox{0.6mm}{$\, >$} \hspace{-3.0mm}\raisebox{-1.5mm}{\em $\sim\,$}}
\newcommand{\vev}[1]{ \left\langle {#1} \right\rangle }
\newcommand{\bra}[1]{ \langle {#1} | }
\newcommand{\ket}[1]{ | {#1} \rangle }
\newcommand{\EV}{ {\rm eV} }
\newcommand{\MeV}{ {\rm MeV} }
\newcommand{\GeV}{ {\rm GeV} }
\newcommand{\TeV}{ {\rm TeV} }
\def\diag{\mathop{\rm diag}\nolimits}
\def\Spin{\mathop{\rm Spin}}
\def\SO{\mathop{\rm SO}}
\def\O{\mathop{\rm O}}
\def\SU{\mathop{\rm SU}}
\def\U{\mathop{\rm U}}
\def\Sp{\mathop{\rm Sp}}
\def\SL{\mathop{\rm SL}}
\def\tr{\mathop{\rm tr}}

\def\IJMP{Int.~J.~Mod.~Phys. }
\def\MPL{Mod.~Phys.~Lett. }
\def\NP{Nucl.~Phys. }
\def\PL{Phys.~Lett. }
\def\PR{Phys.~Rev. }
\def\PRL{Phys.~Rev.~Lett. }
\def\PTP{Prog.~Theor.~Phys. }
\def\ZP{Z.~Phys. }

\def\Z{\mathcal{Z}}
\def\W{\Omega}

\def\stau1{\tilde{\tau}_{1}}
\def\bino{\tilde{B}}
\def\n1{\tilde{\chi}^{0}_{1}}
\def\fsl#1{\setbox0=\hbox{$#1$}                 
   \dimen0=\wd0                                 
   \setbox1=\hbox{/} \dimen1=\wd1               
   \ifdim\dimen0>\dimen1                        
      \rlap{\hbox to \dimen0{\hfil/\hfil}}      
      #1                                        
   \else                                        
      \rlap{\hbox to \dimen1{\hfil$#1$\hfil}}   
      /                                         
   \fi}                                         %


\baselineskip 0.7cm

\begin{titlepage}

\begin{flushright}
UT-11-07\\
TU-880\\
IPMU11-0051\\
KEK-TH-1451\\
\end{flushright}

\vskip 1.35cm
\begin{center}
{\large \bf
Identifying the Origin of Longevity of Metastable Stau at the LHC
}
\vskip 1.2cm
Takumi~Ito$^{1,2}$, Kouhei~Nakaji$^{1,3}$ and Satoshi~Shirai$^{1,4}$
\vskip 0.4cm

{\it
$^1$Department of Physics, University of Tokyo, 
Tokyo 113-0033,
Japan\\
$^2$Department of Physics, Tohoku University, 
Sendai 980-8578,
Japan\\
$^3$IPMU,
University of Tokyo, 
Chiba 277-8586, 
Japan\\
$^4$KEK,
Theory Center,
Tsukuba, 305-0801,
Japan\\
}

\vskip 1.5cm

\abstract{
In the framework of the supersymmetric standard model, the lighter stau often becomes long-lived.
Such longevity of the stau is realized in three well-motivated scenarios:
(A) the stau is the next-to-lightest
supersymmetric particle (NLSP) and the gravitino
is the lightest supersymmetric particle (LSP), 
(B) the stau is the LSP and R-parity is slightly violated, and
(C) the stau is NLSP, the neutralino is the LSP, and the their masses are
degenerate. 
We study the event topology and the decay of the stopping 
stau at the hadron calorimeter
at the LHC, and show that  it is possible to identify the reason why the stau becomes long-lived.
}
\end{center}
\end{titlepage}

\setcounter{page}{1}

\section{Introduction}
A supersymmetric (SUSY) standard  model (SSM) is the most promising candidate of physics beyond the standard model.
In some classes of the SSM, the lighter stau ($\stau1$) is long-lived.
Such long-lived staus have great impact on both cosmology and LHC physics.
Even if such a model is realized, 
the stau cannot be completely stable 
but must have a finite lifetime~\cite{Perl:2001xi}.
If the decay length of the stau is longer than the detector size,
the long-lived massive charged particle will be observed at the LHC~\cite{stau:cmsdiscovery, Nisati:1997gb, polesello_atlmuon, Ambrosanio:2000ik, Ellis:2006vu}.
In this paper, we define the long-lived stau as the stau with the decay length longer than
the detector size.
By using the long-lived stau, it
 is possible to measure the property of the stau, such as the mass, lifetime and so on~\cite{Ambrosanio:2000ik, Ellis:2006vu, Hamaguchi, Ishiwata:2008tp, Kaneko:2008re, Asai:2009ka}.
It is also possible to measure various quantities of the other particles
 using the long-lived stau~\cite{Hinchliffe:1998ys,  Hamaguchi:2004df, Feng:2004yi, Buchmuller:2004rq, Gupta:2007ui, Ibe:2007km, Rajaraman:2007ae, Kitano:2008en, Kitano:2008sa, Feng:2009yq, Biswas:2009zp, Ito:2009xy, Biswas:2009rba, Kitano:2010tt, Biswas:2010cd, Ito:2010xj, Endo:2010ya, Ito:2010un, Asai:2011wy}.

After discovering the stau signals,
it is important how we probe the fundamental physics, using such observation.
To do so, the most important thing is 
 to identify the mechanism which makes the stau meta-stable.
In general, a particle can be (meta-) stable because of
symmetry, weak interaction and/or kinematical reason.
In the framework of the minimal supersymmetric standard model
(MSSM) with the gravitino,
 there are three well-motivated reasons why the stau can be long-lived:
 (A) the stau is the next-to-lightest
supersymmetric particle (NLSP) and the gravitino
is the lightest supersymmetric particle (LSP), 
(B) the stau is the LSP and R-parity is slightly violated, and
(C) the stau is NLSP, the neutralino is the LSP, and the their masses are
degenerate.
Without knowing what makes the stau meta-stable,
it may be impossible to probe the fundamental physics.

In this paper, we discuss the discrimination of the three reasons by studying 
missing energy + stau SUSY events and
the energy deposit from the stau decay which stops at the hadron calorimeter.
We firstly describe the mechanisms which make the lightest stau stable in section~\ref{sec:mechanism}.
In section~\ref{sec:topology}, we show that we can distinguish the case where the stau-longevity attributes to the kinematical reason from the other two cases by using the event topology.
In section~\ref{sec:stopping}, we see that the other two scenarios, the gravitino LSP with the stau
next lightest supersymmetric particle (NLSP) 
and the stau LSP with slightly broken R-parity can be also discriminated by studying the decay of the stau which stops at the hadronic calorimeter. The last section is devoted to conclusions and discussions.

\section{Mechanism of Longevity of Stau}
\label{sec:mechanism}
The lighter stau can be long-lived by several reasons.
Particularly, there are three well-motivated scenarios.
\begin{itemize}
\item[(A)] Weakly interacting LSP:

When there is a weakly interacting LSP and the stau is the NLSP, the stau can be long-lived.
For example, the gravitino and the axino are candidates of such an LSP.
In this paper we discuss the case of gravitino-LSP.
The lifetime of $\stau1$ depends on the gravitino mass~$m_{3/2}$,
\begin{equation}
\tau_{\stau1} \simeq
 \frac{48\pi M_{\rm P}^2 m_{3/2}^2}{m_{\stau1}^5} 
 = 6 \times 10^{-6}~ {\rm sec}
 \left( \frac{m_{\stau1}}{100 \ {\rm GeV}} \right)^{-5} \left( \frac{m_{3/2}}{10 \ {\rm keV}} \right)^{2} ,
\end{equation}
where $M_{\rm P}$ is the reduced Planck mass.
If the gravitino mass is larger than ${\cal O}(1)$~keV, $\stau1$ will be recognized as a long-lived charged particle at the LHC.

\item[(B)] R-parity violation (RPV):

If R-parity is violated, the LSP does not need to be neutral as long as it decays into the standard model sector with a finite lifetime.
Hence, $\stau1$-LSP is possible.
The lifetime of $\stau1$ depends on the size of RPV couplings, and $\stau1$ is long-lived if the couplings are small enough.
$\stau1$ becomes the LSP over the large parameter space of the gravity mediation scenario, and there exist allowed regions where $\stau1$-LSP with small RPV do not contradict to the experiments such as the observation of proton decay~\cite{protondecay} and the cosmological bounds~\cite{bbn}.

\item[(C)] Kinematical suppression:

Even if the R-parity is conserved and the lightest neutralino is the LSP as often assumed in the MSSM, the stau can be long-lived; the decay rate of the stau is highly suppressed if the mass difference between $\stau1$ and $\n1$ is much less than the tauon mass.
It has been discussed that such a scenario is favored in the context of solving the $^7{\rm Li}$ problem~\cite{Jittoh:2010wh}.

\end{itemize}

Depending on the lifetime of $\stau1$, the strategy to probe the origin of the stau longevity is different.
If the typical decay length of $\stau1$ is longer than the detector size, $\stau1$ behaves as if it 
were stable in the LHC experiments, which is the case we study in this paper.
In the case that the most of $\stau1$ decay in the detector in their flight, a study of $\stau1$ kink track becomes more important.
Such a study is found in the recently appeared paper~\cite{Asai:2011wy}.

\section{LHC Event Topology in Long-lived Stau Scenarios}
\label{sec:topology}

Now we start our discussion to probe the origin of the stau longevity at the LHC.
We concentrate on the case where the typical decay length of $\stau1$ is longer than the size of the detector system at the LHC, and in such a case the stau track can be detected by the detector.
Since the stau is a heavy particle, its velocity is relatively small.
The ATLAS detector and the CMS detector can detect such an exotic charged particle
\cite{stau:cmsdiscovery,Aad:2009wy}, relying on the fact that the most of the staus have small velocity.
An existence of the slow-moving charged track with high-$p_T$
is the most characteristic signature of the SUSY event, and standard model backgrounds will be considerably suppressed by requiring the slow-moving track to the signal event.
A part of staus still have a large velocity which may be almost the speed of light.
Such a high-speed stau would not be distinguished from a muon.
We refer the stau track with small velocity to the stau-like track, while the stau-track with large velocity and the ordinary muon track to the muon-like track.
The SUSY events can be categorized by the number of $\stau1$-like tracks: (1)~only one $\stau1$-like track is found, (2)~two $\stau1$-like tracks are found.
This categorization is useful in the following discussions.
We do not consider the third category, no stau-like tracks in the event.

First, consider a scenario that heavy superparticles, such as squarks, always decay into $\stau1$ via cascade decay.
This is the case realized in, e.g., the scenario (A).
Every SUSY event contains two staus from heavy superparticle decay, however, we may find only one stau-like track because we could not distinct the stau-track with large velocity from muons.
Although $\stau1$ mimics a muon when its velocity is so large, such a fake muon generally possesses high-$p_{T}$, $\sim O(100)$~GeV.
Therefore, a high-$p_{T}$ muon will associate to the event with one $\stau1$-like track in this scenario.
A similar discussion would hold for the scenario (B).
If the RPV interactions are weak, heavy superparticles do not decay into the standard model sector, and SUSY cascade decay chains will end up at the LSP, $\stau1$.

In the scenario (C), however,
the above discussion does not hold. In this case, heavy superparticles decay into not only $\stau1$ but $\n1$ with some fraction.
When two SUSY cascade decay chains result in one $\stau1$ and one $\n1$, we will detect only one $\stau1$-like track and, at the same time, observe a large missing transverse energy.
In addition, it is expected that a small number of high-$p_{T}$ muons are associated to the event with one $\stau1$-like track in this scenario.
These signatures are different from the above scenarios, and hence they could be used to probe the origin of the longevity of the stau.

In order to confirm the discussions, we consider three representative scenarios here: (A)~$\stau1$-NLSP in mGMSB~\cite{Dine:1994vc}, (B)~$\stau1$-LSP in mSUGRA with RPV, (C)~mSUGRA with degenerated $\stau1$ and $\n1$ masses.
A mass spectrum of the scenario~(A) is characterized by free parameters: \{$\varLambda$, $M_{\rm mess}$, $N_{5}$, $\tan \beta$, ${\rm sign}(\mu)$\}.
The scenario~(B) and (C) have parameters: \{$m_{0}$, $M_{1/2}$, $A_{0}$, $\tan \beta$, ${\rm sign}(\mu)$\}.\footnote{
Since we pay particular attention to the case with small RPV couplings, RPV effects on the SUSY mass spectrum are negligible.
}
We randomly generate parameter points for each scenarios over the following parameter spaces uniformly:
\begin{eqnarray}
  {\rm (A)} \ :& &\{ 10~{\rm TeV} < \varLambda < 500~{\rm TeV}, \ \varLambda < M_{\rm mess} < 5000~{\rm TeV}, \nonumber \\
  & & \hspace{5mm} 1 \le N_{5} \le 5, \ 1 < \tan \beta < 60, \ {\rm sign}(\mu) = \pm \} \\
  {\rm (B)(C)} \ :& &\{ -3000~{\rm GeV} < m_{0} < 3000~{\rm GeV}, \ 0~{\rm GeV} < M_{1/2} < 3000~{\rm GeV}, \nonumber \\
  & & \hspace{5mm} -3000~{\rm GeV} < A_{0} < 3000~{\rm GeV}, \ 1 < \tan \beta < 60, \ {\rm sign}(\mu) = \pm \}
\end{eqnarray}
We use ISAJET~7.80~\cite{Paige:2003mg} to calculate SUSY mass spectrum and decay modes.
Models are required to have the cross section larger than 0.1~pb in the 14~TeV $pp$-collision so that the large enough number of events will occur at the LHC.
The cross sections of the SUSY events
are calculated by HERWIG~6.510~\cite{Corcella:2000bw,Moretti:2002eu}.
We also require that the stau, winos and colored superparticles (i.e., sup, sdown and gluino) are heavier than 100~GeV, 260~GeV and 450~GeV, respectively.
In addition, we impose a vacuum stability condition, $|A_{\tau}| \tan \beta \sin 2 \theta_{\tau} \lsim 400 \ m_{\stau1}$~\cite{NAKAJI}, where $A_{\tau}$ and $\theta_{\tau}$ are A-parameter and the mixing angle of $\tilde{\tau}$.
For the scenario~(A) and (B), $\stau1$ is the LSP in the MSSM-sector and other superparticle masses are larger than $m_{\stau1} + 1$~GeV.
For the scenario~(C), the same mass condition is imposed to superparticles except for $\n1$; $\n1$ is required to be the LSP while $\stau1$ is the NLSP and their masses are degenerated within 1~GeV.
Then, 2000 parameter points are obtained for each scenario.
After the parameter generation, SUSY events are generated by HERWIG, and then they are passed through the fast detector simulation package PGS~4~\cite{PGS4} with a slight modification to treat stable stau.
$\stau1$ is tagged as $\stau1$-like if its velocity is smaller than $0.9c$, while $\stau1$ with high-velocity is regarded as $\mu$-like.

As we discussed above, a difference of underlying scenarios will be imprinted in a missing energy distribution and a $\mu$-like track distribution associated to the event with one $\stau1$-like track.
Here, we consider two quantities which are useful to extract a nature of the underlying model.
The first one is the difference between the missing $p_{T}$ averaged over
 the events with 1 $\stau1$-like track and that averaged
 over the events with 2 $\stau1$-like tracks,
\begin{equation}
 \Delta \langle \fsl{p}_{T} \rangle \equiv 
 \langle \fsl{p}_{T} \ | {\rm \ 1 \ } \stau1 {\rm -like \ track} \rangle - \langle \fsl{p}_{T} \ | {\rm \ 2 \ }\stau1 {\rm -like \ tracks} \rangle.
\end{equation}
The second one is the average of the number of large-$p_{T}$ muon-like tracks and $\stau1$-like tracks,
\begin{equation}
 \langle N_{\mu, \stau1} \rangle \equiv \langle {\rm the \ number \ of} \ \mu {\rm -like \ tracks \ with \ large-} p_{T} {\rm \ and \ } \stau1 {\rm -like \ tracks} \rangle.
\end{equation}
We use events with at least one $\stau1$-like track.
Then we count up the number of the $\stau1$-like tracks and the $\mu$-like tracks with $p_{T} > m_{\stau1}$.\footnote
{
The mass of $\stau1$ is directly measurable using its track information~\cite{stau:cmsdiscovery,Aad:2009wy}.
Hence we expect that the mass determination is accurate enough by using a small number of the tracks~\cite{stau:cmsdiscovery}.
}
\footnote
{
The result does not change drastically as long as we adopt the $p_{T}>{\cal O}(m_{\stau1})$.
}
Notice that these quantities well describe the underlying model when the staus produced from heavy superparticle cascades.
Therefore we use the event in which the leading jet has $p_{T} > 80$~GeV.
In addition, we require that at least one of the following conditions is satisfied:
 (1) a second jet with $p_{T} > 60$~GeV,
 (2) an electron with $p_{T} > 30$~GeV,
 (3) a $\tau$-tagged jet with $p_{T} > 40$~GeV.

\begin{figure}[t]
  \centerline{\epsfxsize=0.85\textwidth\epsfbox{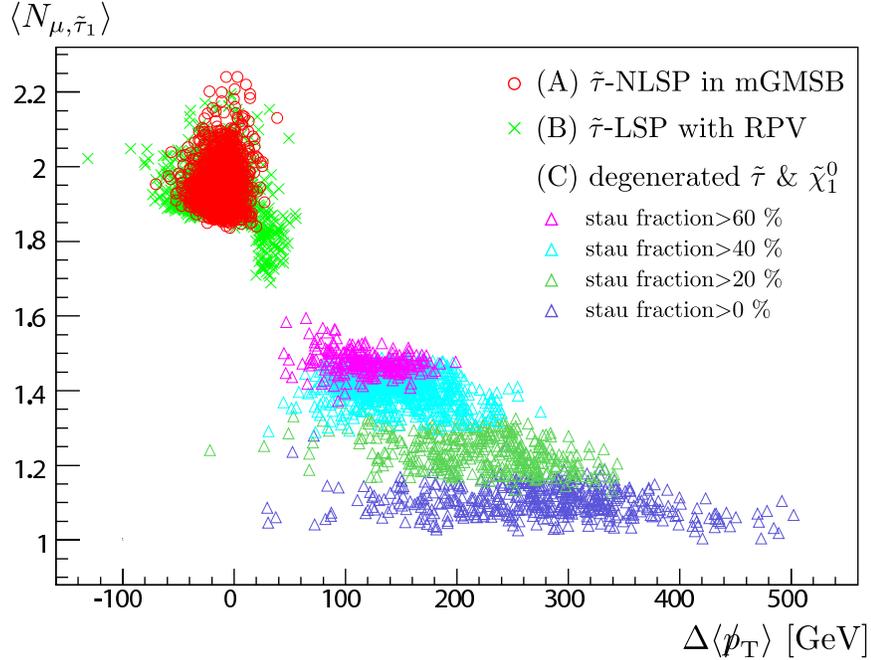}}
  \caption{\small Distribution of $\Delta \langle \fsl{p}_{T} \rangle$ and $\langle N_{\mu, \stau1} \rangle$ in three scenarios. The stau fraction is at most $\sim 80\% $ in the scenario~(C). On the contrary, the fraction is 100\% in the other scenarios.}
  \label{fig:mpt_nmu}
\end{figure}

In Fig.~\ref{fig:mpt_nmu}, we show the distribution of these quantities in three scenarios.
In scenario~(A) and (B), two staus arise in the final state.
Even if we cannot distinguish a high-velocity stau from muons, the resulting $\mu$-like track would have large $p_{T}$.
Hence we expect $\langle N_{\mu, \stau1} \rangle \simeq 2$.
In these cases, sources of the missing energy are mainly neutrinos and miss-measurement of jet energies.
If a correlation between the number of $\stau1$-like tracks and the missing energy sources is weak, we expect $\Delta \langle \fsl{p}_{T} \rangle \simeq 0$.
On the other hand, the final state would contain not only $\stau1$ but $\n1$ in scenario~(C).
Therefore, we expect that $\langle N_{\mu, \stau1} \rangle < 2$ and $\Delta \langle \fsl{p}_{T} \rangle > 0$ in this scenario.
These expectations are indeed true as one can see from the figure, and the scenario~(C) would be discerned from the other two scenarios. 

For the discrimination of the scenario~(C) from the other scenarios,
roughly speaking,  
we need to determine the $\langle N_{\mu, \stau1} \rangle$ and $\Delta \langle \fsl{p}_{T} \rangle$ with the accuracies of 0.2 and 50~GeV, respectively.
We estimate the sizes of the statistical fluctuations.
With 100 signal events, $\delta ( \Delta \langle \fsl{p}_{T} \rangle ) \simeq \pm 50$~GeV and $\delta ( \langle N_{\mu, \stau1} \rangle) \simeq \pm 0.1$.
In the figure, we indicate a fraction of the stau appeared in the final state
at the parton level.
Both $\Delta \langle \fsl{p}_{T} \rangle$ and $\langle N_{\mu, \stau1} \rangle$ are strongly correlated to the stau fraction.
It will be possible to determine the stau fraction when the large number of signal events are observed.

\section{Decay of Stopping stau at the Hadron Calorimeter}
\label{sec:stopping}

At the LHC, the stau with small velocity ($\beta \gamma \lsim 0.45$) 
lose their kinetic energy by ionization and stops at the hadron calorimeter \cite{Asai:2009ka}. 
The ATLAS
detector can identify the stopping-stau events and can
observe the energy deposit of 
hadrons produced by the decay of such stopping 
staus with a wide range of the lifetime 
${\cal O}(0.1) - {\cal O}(10^{10})$ sec at the hadron calorimeter \cite{Asai:2009ka}. 
Non hadronic objects from the stau are difficult to be observed.

In the previous section,
we found that by observing the LHC event topology,
we can discriminate the degenerated mass scenario from the
other two scenarios: (A) the gravitino LSP
with $\stau1$-NLSP and (B) $\stau1$-LSP with RPV.
In this section, 
we show that
by studying the 
decay of stopping staus at the hadronic calorimeter,
we can distinguish between the scenario (A) and the scenario (B) which 
are not distinguished by the LHC event topology.

We use following two observable quantities
in order to distinguish the two scenarios. The first one is $P^{\rm had}_{\tilde{\tau}}$ defined by
\begin{eqnarray}
P^{\rm had}_{\tilde{\tau}}=
\frac{\rm the~number~of~events~where~hadrons~from~the~stau~decay~are~observed}
{\rm the~number~of~the~stopping~staus}.
\end{eqnarray}
The second one is the energy distribution of hadrons produced by the stau decay.
Here we assume that all of the staus decay within the time 
that the detector is prepared to observe the stau decay.\footnote{
For detailed discussion about how to observe the decay of the stopping stau with
various lifetime, see \cite{Asai:2009ka}.
}
We also assume that we can observe all of the energy deposit of
hadrons produced from the stau decay.\footnote{
We 
include all of the energy of quarks into the energy of hadrons though
quarks may radiate some non-hadronic particles. The energy of 
$\pi^0$ produced by tau-decay is also included in the energy of hadrons.
}
Under these assumptions,
 $P^{\rm had}_{\tilde{\tau}}$ is equal to the
 branching fraction of the stau into at least one hadron.
We use the programs Madgraph4.0 \cite{Madgraph} 
and TAUOLA2.6 \cite{TAUOLA} to simulate the energy distribution.

\begin{figure}
\begin{center}
\includegraphics[width=10cm]{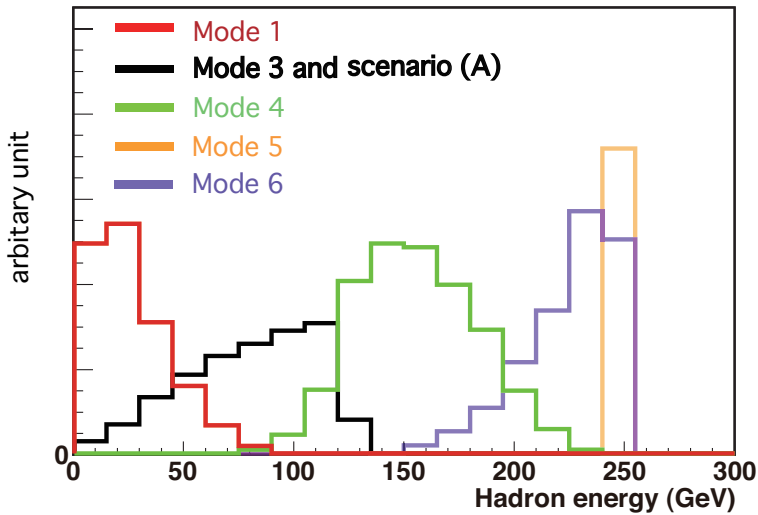}
\end{center}
\caption{Energy distribution of hadrons in each scenario for $m_{\stau1}=250~{\rm GeV}$.
The normalization is arbitrary. Mode 1 and Mode 3-6 are the cases with
RPV decay. (See Table \ref{Rparitydecay}.)
}
\label{energydistribution}
\end{figure}

\begin{figure}
\begin{center}
\includegraphics[width=10cm]{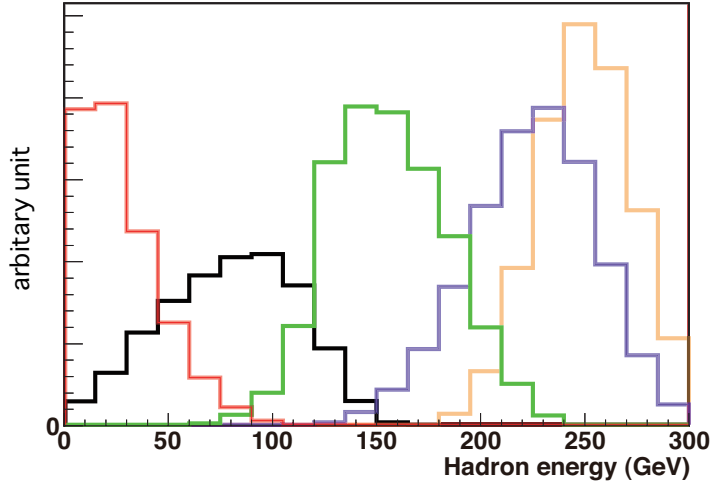}
\end{center}
\caption{Same as Fig.\ref{energydistribution} but with the energy resolution  
$\Delta E /E=150 \% / \sqrt{E/{\rm GeV}}$.}
\label{smeared}
\end{figure}

In the case of (A), a stau decays into a gravitino as $\stau1\rightarrow \tau \tilde{G}$
and the tau immediately decays leptonically or hadronically. Because the event with no hadrons 
can not be triggered at the detector, we can observe only the events where the tau decays
hadronically. Therefore,  $P^{\rm had}_{\tilde{\tau}}$ in this scenario
is equal to the hadronic branching fraction of tau $P_{\tau}$ ($\simeq0.65$). 
The energy distribution of hadrons is shown by the black line in
Fig.~\ref{energydistribution}.
In the figure, 
we assume that the stau is purely right handed, and  
the stau mass
is set to be $250~{\rm GeV}$.
 We also show the energy
distribution with the detector resolution $\Delta E /E=150 \% / \sqrt{E/{\rm GeV}}$
in Fig.~\ref{smeared}.

On the other hand, in the case of (B), many types of the decay
modes are possible depending on the coupling constant of R-parity violating term. 
Here we
concentrate on the trilinear R-parity violating terms in the superpotential,

\begin{eqnarray}
\label{Rviolation}
W_{R\hspace{-0.40em}/}=\frac{1}{2}\lambda_{ijk}L_iL_j\bar{E}_k
+\lambda^{\prime}_{ijk}L_{i}Q_{j}\bar{D}_k
+\frac{1}{2}\lambda^{\prime\prime}_{ijk}\bar{U}_{i}\bar{D}_{j}\bar{D}_k,
\end{eqnarray}
where $L_i$ and $Q_i$ are superfield of SU(2) doublet of left-handed leptons and left handed quarks respectively. $E_i$, $U_i$ and $D_i$ denote the right-handed 
superfield of  leptons, up-type quarks and down-type quarks respectively. 
The decay pattern of the stau by the trilinear R-parity violating coupling has been already
discussed in the other context \cite{dreiner}.
We summarize
the decay modes and 
the hadronic branching ratio of staus when one R-parity violating coupling dominates in Table \ref{Rparitydecay}. 
We show the energy distributions corresponding to each modes 
in Fig.~\ref{energydistribution} where 
we assume that the stau is purely right handed.
The smeared energy distributions are shown in 
Fig.~\ref{smeared}.
\begin{table}
\begin{center}
\begin{tabular}{|c|c|c|c|} 
\hline
Mode&coupling&decay modes& hadronic branch  \\ \hline
Mode 1&$\lambda_{121}=-\lambda_{211}$&$\tau^{+} e^{+}_i e^{-}_j \nu_k$&$P_{\tau}$ \\ 
&$\lambda_{122}=-\lambda_{212}$&$\tau^{+} e^{+}_i e^{-}_j \bar{\nu_k}$& \\ \hline
&$\lambda_{131}=-\lambda_{311}$&& \\ 
&$\lambda_{132}=-\lambda_{312}$&$e_i^{+} \nu_{j}$& \\
Mode 2&$\lambda_{231}=-\lambda_{321}$&$e_i^{+} \bar{\nu_j}$& $\ll 1$\\ 
&$\lambda_{232}=-\lambda_{322}$&& \\
&$\lambda_{123}=-\lambda_{213}$&& \\ 
\hline
Mode 3&$\lambda_{133}=-\lambda_{313}$&$e_i^+$$\nu_{\tau}$ &$P_{\tau}/2$ 
(purely right handed)
\\
&$\lambda_{233}=-\lambda_{323}$&$e_i^+$$\bar{\nu_{\tau}}$ &$\wr$\\
&&$\tau^+ \nu_i$&$P_{\tau}$ (purely left handed) \\
&&$\tau^+ \bar{\nu_i}$&\\  \hline
Mode 4&$\lambda^{\prime}_{1lm}$&$\tau^+ \bar{u_i}d_j e_k^+$&\\
&$\lambda^{\prime}_{2lm}$&$\tau^+ u_i \bar{d_j} e_k^-$&1\\
&&$\tau^+ \bar{d_i} d_j \bar{\nu_k} $&\\
&&$\tau^+ \bar{d_i} d_j \nu_k $&\\ \hline
Mode 5&$\lambda^{\prime}_{3lm}$&$u_j \bar{d_k}$&$1$\\  \hline
Mode 6&$\lambda^{\prime \prime}_{lmn}$&$\tau^+ u_i d_j d_k$&$1$\\
&&$\tau^+ \bar{u_i} \bar{d_j} \bar{d_k}$&\\ 
\hline
\hline
\multicolumn{2}{|c|}{Scenario (A)}&$\tau^+ \tilde{G}$&$P_{\tau}$\\
\hline
\end{tabular}
\end{center}
\caption{Decay mode and 
hadronic branching ratio of $\stau1^+$
when one R-parity violating coupling dominates. 
We also show 
those of the scenario (A) for the reference.
$(e_1, e_2)=(e,\mu)$, $( \nu_1, \nu_2)=(\nu_e, \nu_{\mu})$,
$(u_1,u_2,u_3)=(u,c,t)$ and $(d_1,d_2,d_3)=(d,s,b)$.  }
\label{Rparitydecay}
\end{table}

We find that
we can distinguish the scenario (A) from Mode 2--6 by observing $P^{\rm had}_{\tilde{\tau}}$.
When ${\cal O}(10)$ of stau decays are observed,
 we can discriminate scenario (A) from Mode 2--6 in $3$-$\sigma$ confidence
 level. 
It is also found that we can discriminate the scenario (A) from Mode 1
by observing the energy distributions of hadrons. 
We check that by using the maximum likelihood analysis to 
the smeared energy distributions of scenario (A) and Mode 1 in scenario (B),
the discrimination of
the two energy distributions in 3-$\sigma$ confidence level is
possible by observing 
${\cal O}(10)$ of stau decays. Thus, by observing ${\cal O}(10)$ of stau decays,
we can distinguish the scenario (A) with (B) in $3$-$\sigma$ confidence level.
It is also
checked that even when the several modes contribute, we can distinguish
the scenario (A) with (B) by using the hadronic branching fraction and the
energy distributions. 
In the realistic detector system, 
only the hadrons with the energy more than $20~{\rm GeV}$ can be observed,
 however the conclusion does not change by the energy cut.

In this paper, we do not consider the bilinear term $\epsilon_i L_i H_u$
in RPV case 
where $H_u$ is the superfield of up-type Higgs. 
In the scenario with the bilinear couplings, 
there are many types of the decay mode 
such as $\stau1 \rightarrow \tau \bar{\nu}$,
$\stau1 \rightarrow t \bar{b}$ and $\stau1 \rightarrow WZ$.  
We can discriminate the two scenarios by using the same way
in the large parameter region.
In some cases, unfortunately,
a stau mainly decays by the process $\stau1 \rightarrow \tau \bar{\nu}$ or
$\stau1 \rightarrow \tau \nu$, and therefore 
the hadronic branching ratio and the energy distribution of stau is almost the same as
the stau NLSP case. In such cases, it is difficult to distinguish the scenario (A) from 
the scenario (B).

The higher dimensional RPV operators are also possible.
With the higher dimensional RPV operator,
we expect that 
by comparing the energy distribution, it is possible to discriminate the 
two scenarios because it is expected that
the decay using higher dimensional operator tend to produce more than
2 particles. 
 The study of the case with the bilinear terms and
the higher dimensional operators are for future work.

\section{Conclusion and Discussion}

In this paper, we have discussed how to prove the origin of the stau-longevity at the LHC experiments.
There exist three distinct mechanisms which make the lighter stau long-lived:
weak interaction between $\stau1$-NLSP and LSP,
$\stau1$-LSP with small R-parity violation and 
suppressed $\stau1$ decay into the LSP due to the mass degeneracy.

If the masses of the $\stau1$-NLSP and $\chi^{0}_{1}$-LSP are sufficiently
 degenerated, the stau does not promptly decay into the neutralino due to the kinematical suppression.
In such a scenario, heavy superparticles produced at the LHC will decay into not only the stau but the neutralino.
This feature can be probed via SUSY event topology.
In particular, the missing energy distribution and the number of high-$p_{T}$ muons provide useful information.
They have been investigated in section \ref{sec:topology}, and 
it is shown to be possible to check the degeneracy between the stau and the neutralino at the early stage of the LHC experiment.

The decay of the stopping stau at the calorimeter provides another important information on the origin of the longevity as discussed in section \ref{sec:stopping}.
If the stau is the NLSP and there is the weakly interacting LSP, e.g., gravitino or axino, $\stau1$ will produce $\tau$ in its decay into the neutral LSP.
In the case of the stau-LSP with $R$-parity violation, there are various stau decay modes depending on the RPV couplings but they generally posses different final states from the $\stau1$-NLSP case.
A hadronic branching fraction of the stau decay can be used to discriminate the $\stau1$-NLSP case from some of RPV cases.
Also, an energy distribution of hadrons from stau decay is useful.
Using these informations, we could discriminate the cases where RPV is induced by
the trilinear coupling 
 from the $\stau1$-NLSP scenarios.

In this paper, we assume that 
all the decays of trapped staus
occur in operation time of the detector, 
and that
the detection efficiency for the hadronic decay of the stau trapped in the calorimeter is 100 \%.
In practice, there need be some corrections to estimate the hadronic branching fraction
because some of the decays occur in the dead time of the detector, depending on the lifetime of the stau.
Although it would be also necessary to consider the detection inefficiency,
it is expected that such effects do not change the present discussion drastically.
Detailed treatments of such effects are beyond the scope of this paper.

Let us comment on the application range for the present method.
The method of the event topology requires ${\cal O}(100)$ events
in order to discriminate the models.
The method of the measurement of the stopped stau decay in the calorimeter needs ${\cal O}(10)$ stopping events. Such stopping events are typically ${\cal O}(1)~\%$ of all SUSY 
events \cite{Asai:2009ka}. 
 Therefore, the present method is applicable to the case that $m_{\tilde{W}} \lsim $ 500 GeV or $m_{\tilde{g},\tilde{q}}\lsim $ 1500 GeV for the integrated luminosity 10 ${\rm fb}^{-1}$ at
$\sqrt{s}=14$ TeV at the LHC.

\section*{Acknowledgements}
We would like to thank K.~Hamaguchi, S.~Asai, S.~Iwamoto, T.~Moroi and T.~Yanagida for useful comments
and helpful discussions.
This work is supported in part by JSPS
Research Fellowships for Young Scientists and by
World Premier International Research Center Initiative, MEXT, Japan.


\begin{thebibliography}{99}
\bibitem{Perl:2001xi}
  M.~L.~Perl, P.~C.~Kim, V.~Halyo, E.~R.~Lee, I.~T.~Lee, D.~Loomba and K.~S.~Lackner,
  Int.\ J.\ Mod.\ Phys.\  A {\bf 16}, 2137 (2001)

\bibitem{stau:cmsdiscovery}
  The CMS Collaboration, CMS-PAS-EXO-08-003;
  The CMS Collaboration, CMS-NOTE-2010-008.

\bibitem{Nisati:1997gb}
  A.~Nisati, S.~Petrarca and G.~Salvini,
  Mod.\ Phys.\ Lett.\  A {\bf 12}, 2213 (1997).

\bibitem{polesello_atlmuon}
G. Polesello and A. Rimoldi, ATL-MUON-99-006.

\bibitem{Ambrosanio:2000ik}
  S.~Ambrosanio, B.~Mele, S.~Petrarca, G.~Polesello and A.~Rimoldi,
  JHEP {\bf 0101}, 014 (2001).

\bibitem{Ellis:2006vu}
  J.~R.~Ellis, A.~R.~Raklev and O.~K.~Oye,
  JHEP {\bf 0610}, 061 (2006).


\bibitem{Hamaguchi}
  K.~Hamaguchi, M.~M.~Nojiri and A.~de Roeck,
  JHEP {\bf 0703}, 046 (2007).


\bibitem{Ishiwata:2008tp}
  K.~Ishiwata, T.~Ito and T.~Moroi,
  Phys.\ Lett.\  B {\bf 669}, 28 (2008).

\bibitem{Kaneko:2008re}
  S.~Kaneko, J.~Sato, T.~Shimomura, O.~Vives and M.~Yamanaka,
  Phys.\ Rev.\  D {\bf 78}, 116013 (2008).

\bibitem{Asai:2009ka}
  S.~Asai, K.~Hamaguchi and S.~Shirai,
  Phys.\ Rev.\ Lett.\  {\bf 103}, 141803 (2009).


\bibitem{Hinchliffe:1998ys}
  I.~Hinchliffe and F.~E.~Paige,
  Phys.\ Rev.\  D {\bf 60}, 095002 (1999).

\bibitem{Hamaguchi:2004df}
  K.~Hamaguchi, Y.~Kuno, T.~Nakaya and M.~M.~Nojiri,
  Phys.\ Rev.\  D {\bf 70}, 115007 (2004).

\bibitem{Feng:2004yi}
  J.~L.~Feng and B.~T.~Smith,
  Phys.\ Rev.\  D {\bf 71}, 015004 (2005)
  [Erratum-ibid.\  D {\bf 71}, 019904 (2005)].

\bibitem{Buchmuller:2004rq}
  W.~Buchmuller, K.~Hamaguchi, M.~Ratz and T.~Yanagida,
  Phys.\ Lett.\  B {\bf 588}, 90 (2004).

\bibitem{Gupta:2007ui}
  S.~K.~Gupta, B.~Mukhopadhyaya and S.~K.~Rai,
  Phys.\ Rev.\  D {\bf 75}, 075007 (2007).


\bibitem{Ibe:2007km}
  M.~Ibe and R.~Kitano,
  JHEP {\bf 0708}, 016 (2007).

\bibitem{Rajaraman:2007ae}
  A.~Rajaraman and B.~T.~Smith,
  Phys.\ Rev.\  D {\bf 76}, 115004 (2007).

\bibitem{Kitano:2008en}
  R.~Kitano,
  JHEP {\bf 0803}, 023 (2008).


\bibitem{Kitano:2008sa}
  R.~Kitano,
  JHEP {\bf 0811}, 045 (2008).

\bibitem{Feng:2009yq}
  J.~L.~Feng, S.~T.~French, C.~G.~Lester, Y.~Nir and Y.~Shadmi,
  Phys.\ Rev.\  D {\bf 80}, 114004 (2009);
  J.~L.~Feng {\it et al.},
  JHEP {\bf 1001}, 047 (2010).

\bibitem{Biswas:2009zp}
  S.~Biswas and B.~Mukhopadhyaya,
  Phys.\ Rev.\  D {\bf 79}, 115009 (2009).

\bibitem{Ito:2009xy}
  T.~Ito, R.~Kitano and T.~Moroi,
  JHEP {\bf 1004}, 017 (2010).

\bibitem{Biswas:2009rba}
  S.~Biswas and B.~Mukhopadhyaya,
  Phys.\ Rev.\  D {\bf 81}, 015003 (2010).

\bibitem{Kitano:2010tt}
  R.~Kitano and M.~Nakamura,
  Phys.\ Rev.\  D {\bf 82}, 035007 (2010).

\bibitem{Biswas:2010cd}
  S.~Biswas,
  Phys.\ Rev.\  D {\bf 82}, 075020 (2010).

\bibitem{Ito:2010xj}
  T.~Ito and T.~Moroi,
  Phys.\ Lett.\  {\bf B694}, 349-354 (2011).

\bibitem{Endo:2010ya}
  M.~Endo, K.~Hamaguchi and K.~Nakaji,
  JHEP {\bf 1011}, 004 (2010).

\bibitem{Ito:2010un}
  T.~Ito,
  arXiv:1012.1318 [hep-ph].  

\bibitem{Asai:2011wy}
  S.~Asai, Y.~Azuma, M.~Endo, K.~Hamaguchi, S.~Iwamoto,
  arXiv:1103.1881 [hep-ph].
  
  
  

\bibitem{protondecay}
 C. Amsler, et al. Review of particle physics. Phys. Lett., B667, 1 (2008). and 2009 partial
update at http://pdg.lbl.gov/.  pp. 2, 5, 11, 12, 28, 68, 73, and 75.

\bibitem{bbn}
 M.~Kawasaki, K.~Kohri, T.~Moroi and A.~Yotsuyanagi,
 Phys.\ Rev.\  D {\bf 78}, 065011 (2008).

\bibitem{Jittoh:2010wh}
  T.~Jittoh, K.~Kohri, M.~Koike, J.~Sato, T.~Shimomura and M.~Yamanaka,
  Phys.\ Rev.\  D {\bf 82}, 115030 (2010).


\bibitem{Dine:1994vc}
  M.~Dine, A.~E.~Nelson and Y.~Shirman,
  Phys.\ Rev.\  D {\bf 51}, 1362 (1995);
  M.~Dine, A.~E.~Nelson, Y.~Nir and Y.~Shirman,
  Phys.\ Rev.\  D {\bf 53}, 2658 (1996).


\bibitem{Paige:2003mg}
  F.~E.~Paige, S.~D.~Protopopescu, H.~Baer and X.~Tata,
  arXiv:hep-ph/0312045.

\bibitem{Corcella:2000bw}
  G.~Corcella {\it et al.},
  JHEP {\bf 0101}, 010 (2001);
%
  arXiv:hep-ph/0210213.

\bibitem{Moretti:2002eu}
  S.~Moretti, K.~Odagiri, P.~Richardson, M.~H.~Seymour and B.~R.~Webber,
  JHEP {\bf 0204}, 028 (2002).

\bibitem{NAKAJI}
 M. Endo, K. Hamaguchi and K. Nakaji, in preparation.

\bibitem{PGS4} 
  For information on Pretty Good Simulation of high energy
  collisions (PGS4), see
  {\verb$http://www.physics.ucdavis.edu/%7Econway/research/research.html$}.

\bibitem{Aad:2009wy}
  G.~Aad {\it et al.}  [The ATLAS Collaboration],
  arXiv:0901.0512 [hep-ex].



\bibitem{Madgraph}
  F.~Maltoni and T.~Stelzer,
  JHEP {\bf 0302}, 027 (2003).

\bibitem{TAUOLA}
  S.~Jadach, J.~H.~Kuhn, Z.~Was,
  Comput.\ Phys.\ Commun.\  {\bf 64}, 275-299 (1990).

\bibitem{dreiner}
  K.~Desch, S.~Fleischmann, P.~Wienemann, H.~K.~Dreiner and S.~Grab,
  Phys.\ Rev.\  D {\bf 83}, 015013 (2011).
\end{thebibliography}
\end{document}